\documentclass[a4paper,reqno,12pt,final]{amsart}
\usepackage{amsmath,amsthm,amssymb,amscd,euscript,bbold}
\usepackage{array}
\usepackage[T1]{fontenc}
\usepackage[latin1]{inputenc}
\newcommand{\RR}{\mathbb{R}}

\newcommand{\Spin}{\mathrm{Spin}}

\newcommand{\SU}{\mathrm{SU}}

\renewcommand{\Sp}{\mathrm{Sp}}

\newcommand{\D}{\mathsf{D}}
\newcommand{\M}{\mathsf{M}}

\newcommand{\EE}{\mathbb{E}}

\newcommand{\half}{\tfrac{1}{2}}

\renewcommand{\d}{\partial}

\theoremstyle{plain}

\theoremstyle{definition}

\theoremstyle{remark}

\begin{document}

\title{More Ricci-flat branes}
\author[JM Figueroa-O'Farrill]{José Miguel Figueroa-O'Farrill}
\address{The Erwin Schrödinger International Institute for Mathematical
  Physics, Boltzmanngasse 9, A-1090 Wien, Austria}
\curraddr{Department of Mathematics and Statistics, University of
  Edinburgh, King's Buildings (JCMB), Edinburgh EH9
  3JZ, Scotland}
\email{jmf@maths.ed.ac.uk}
\date{\today}
\begin{abstract}
  Certain supergravity solutions (including domain walls and the
  magnetic fivebrane) have recently been generalised by Brecher and
  Perry by relaxing the condition that the brane worldvolume be flat.
  In this way they obtain examples in which the brane worldvolume is a
  static spacetime admitting parallel spinors.  In this note we simply
  point out that the restriction to static spacetimes is unnecessary,
  and in this way exhibit solutions where the brane worldvolume is an
  indecomposable Ricci-flat lorentzian manifold admitting parallel
  spinors.  We discuss more Ricci-flat fivebranes and domain walls, as
  well as new Ricci-flat $\D3$-branes.
\end{abstract}
\maketitle

\tableofcontents

\section{Introduction}

Classical solutions to supergravity theories, particularly those which
preserve some of the supersymmetry, have played an enormously important
role in many of the recent developments in nonperturbative string theory
as well as in establishing new correspondences between gravity and gauge
theory.  Notwithstanding the vigorous effort that has been made in the
last several years and the rapid progress in understanding the geometry
of these solutions, one is nevertheless left with the feeling that we
have merely scratched the surface of what appears to be a very rich
moduli space.

Roughly speaking the metric of a typical $p$-brane solution is a warped
product of two spacetimes: the worldvolume of the brane (traditionally
taken to be $(p+1)$-dimensional Minkowski spacetime) and the transverse
space to the brane.  In the elementary brane solutions, the transverse
space is taken to be euclidean space; but by now many solutions have
been obtained where the transverse geometry is more interesting.  Not
all of these solutions correspond to localised branes, but many which do
can be interpreted as branes localised at conical singularities in
manifolds of special holonomy, the reduction of the holonomy being a
familiar phenomenon in supersymmetry.

In light of the body of work on $\D$-branes in curved manifolds (see,
e.g., \cite{FSDNW} and references therein), where brane worldvolumes
are generically curved, the restriction to flat brane worldvolumes
seems unnatural, and in recent work \cite{BP} Brecher and Perry set
out to construct examples of supersymmetric brane solutions where the
brane worldvolume is taken to be a curved manifold.

Their analysis shows that for the supergravity equations of motion to be
obeyed, the brane worldvolume must be Ricci-flat, whereas supersymmetry
dictates that it should in addition admit parallel (i.e., covariantly
constant) spinors.  Up to dimension three, Ricci-flatness implies
flatness, but for $p$-branes with $p\geq 3$, one should be able to write
down solutions with a curved brane worldvolume.  In \cite{BP} the
authors discuss mainly domain walls (i.e., branes whose worldvolumes
have codimension $1$) and the magnetic fivebrane.  Furthermore they use
the known classification of \emph{riemannian} holonomy groups to attempt
to classify the possible brane worldvolumes.

This analysis contains the tacit assumption that the brane worldvolume
$B$ of the $p$-brane is a \emph{static} spacetime; that is, its metric
is of the form:
\begin{equation*}
  ds^2(B) = -dt^2 + ds^2(X)~,
\end{equation*}
where $X$ is a $p$-dimensional riemannian manifold admitting parallel
spinors.  Such a metric is automatically Ricci-flat, and the
(restricted) holonomy group $H$ of $X$ is constrained to be one of the
groups in Table~\ref{tab:static} (see, e.g., \cite{Wang}).

\begin{table}[h!]
\centering
\renewcommand{\arraystretch}{1.3}
\begin{tabular}{|>{$}r<{$}|>{$}c<{$}|}\hline
  p & H \subset \Spin(p) \\
  \hline\hline
   9 & \Spin(7) \times \{1\}\\[3pt]
   8 & \Spin(7) \\
   7 & G_2 \\[3pt]
   6 & \SU(3) \\[3pt]
   5 & \Sp(1) \times \{1\}\\[3pt]
   4 & \Sp(1) \\[3pt]
   \leq3 & \{1\} \\ \hline
\end{tabular}
\vspace{8pt}
\caption{Restricted holonomy groups of the $p$-dimensional manifold $X$.
  Only maximal subgroups are shown.  For example in dimension $8$ one
  can have subgroups of $\Spin(7)$: $\SU(4)\supset Sp(2)\supset \dots$,
  and so on.}
\label{tab:static}
\end{table}

The restriction to static spacetimes is however unnecessary from a
geometrical point of view, and the point of this note is to exhibit a
roughly equally large class of solutions where the brane worldvolume is
an indecomposable Ricci-flat lorentzian spacetime admitting parallel
spinors.  As explained, for example in \cite{JMWaves}, a lorentzian
manifold admitting parallel spinors need not be Ricci-flat; so it is not
enough to restrict the holonomy in order to satisfy the equations of
motion.  The results described below are a direct corollary of the
methods, if not the results, of the paper \cite{JMWaves}, to which the
reader is referred for details not covered here.  For the present
purposes we restrict ourselves to partially reproducing in
Table~\ref{tab:spinisotropy}, one of the tables in that paper,
containing a (conjecturally complete) list of (restricted) holonomy
groups of indecomposable lorentzian manifolds admitting parallel
spinors.  For an explanation of how to understand these subgroups in
terms of Lorentz transformations, as well as for the notion of
indecomposability, we refer the reader to the Appendix and Section~2.3
of \cite{JMWaves}, respectively.

\begin{table}[h!]
\centering
\renewcommand{\arraystretch}{1.3}
\begin{tabular}{|>{$}r<{$}|>{$}c<{$}|}\hline
  d & H \subset \Spin(d-1,1) \\
  \hline\hline
  9 & G_2 \ltimes \RR^7 \\[3pt]
  8 & \SU(3) \ltimes \RR^6 \\[3pt]
  7 & \left(\Sp(1) \ltimes \RR^4\right) \times \RR \\[3pt]
  6 & \Sp(1) \ltimes \RR^4 \\[3pt]
  \leq 5 & \RR^{d-2} \\
  \hline
\end{tabular}
\vspace{8pt}
\caption{Restricted holonomy groups of $d$-dimensional indecomposable
lorentzian manifolds admitting parallel spinors.  Only maximal
subgroups are shown.}
\label{tab:spinisotropy}
\end{table}

This note is organised as follows.  In the next two sections we apply
the methods of \cite{JMWaves} to two of the examples of branes discussed
by Brecher and Perry: the magnetic fivebrane of eleven-dimensional
supergravity in Section~2 and the $\D8$-brane in type IIA supergravity
in Section~3, constructing new solutions in each case.  Finally in
Section~4 we discuss new Ricci-flat $\D3$-branes.

After this note had been circulated, we became aware of the paper
\cite{Janssen} where branes with curved worldvolumes are also
considered.

\subsubsection*{Acknowledgements}
This note was written while I was participating in the programme
\emph{Holonomy groups in differential geomety} at the Erwin Schrödinger
International Institute for Mathematical Physics.  It is a pleasure to
thank the organisers, particularly Krzysztof Galicki for the invitation
and Dmitri Alekseevsky for the opportunity to speak on this subject, as
well as the other participants for providing such a stimulating
atmosphere.  Last but not least, I would like to acknowledge the
financial support of the Royal Society and the ESI during this visit.

\section{More Ricci-flat fivebranes}

The metric for the magnetic fivebrane of eleven dimensional supergravity 
is obtained by warping the flat metric on the minkowskian worldvolume of 
the fivebrane with the flat metric of the five dimensional transverse
Euclidean space.  The background describing a number of parallel
fivebranes is given by \cite{Guven}:
\begin{align}\label{eq:M5}
ds^2 &= H^{-\frac13}\, ds^2(\EE^{5,1}) + H^{\frac23} ds^2(\EE^5)\\
F &= \pm 3 \star_5 dH\notag~,
\end{align}
where $ds^2(\EE^{5,1})$ is the metric on the six-dimensional Minkowskian
worldvolume $\EE^{5,1}$ of the branes, $ds^2(\EE^5)$ and $\star_5$
are the metric and Hodge operator on the five-dimensional euclidean
space $\EE^5$ transverse to the branes, and $H$ is the following
harmonic function on $\EE^5$:
\begin{equation}\label{eq:H(r)5}
H(r) = 1 + \frac{a^3}{r^3}~; \qquad a^3 \equiv \pi N \ell_p^3
\end{equation}
where $r$ is the transverse radial coordinate on $\EE^5$, $N$ is an
positive integer and $\ell_p$ is the eleven-dimensional Planck
constant.  We have fixed the constant in $H(r)$ so that
$\lim_{r\to\infty} H(r) = 1$.  This solution then corresponds to $N$
coincident fivebranes at $r=0$.  As is well known, this solution
preserves a fraction $\nu=\half$ of the supersymmetry.

The analysis of \cite{BP} shows that the above expression remains a
supersymmetric solution provided that $ds^2(\EE^{5,1})$ is replaced by a
Ricci-flat Lorentzian metric admitting parallel spinors.  One
possibility, discussed in \cite{BP}, is to consider a product spacetime
$\EE^{1,1} \times K$, with $K$ a hyperkähler four-dimensional manifold.
Such a metric is automatically Ricci-flat and since the solution is
half-flat, it preserves a fraction $\nu = \tfrac14$ of the
supersymmetry.

Alternatively, one can take an indecomposable Ricci-flat lorentzian
metric admitting parallel spinors.  From Table~\ref{tab:spinisotropy}
one reads off that the (restricted) holonomy of such a metric must be
contained in a subgroup $\Sp(1)\ltimes \RR^4 \subset \Spin(5,1)$.  In
fact, indecomposability means that there are only two possible subgroups 
$\RR^4$ and $\Sp(1)\ltimes \RR^4$.  As discussed in \cite{JMWaves}, the
former subgroup preserves a fraction $\nu=\tfrac14$  of the
supersymmetry, whereas the latter preserves a fraction $\nu=\tfrac18$.

Since for these groups, the holonomy reduction does not guarantee
Ricci-flatness, which is necessary to satisfy the equations of motion,
we have to make sure that Ricci-flat metrics with this holonomy do
indeed exist.

The most general local metric in six dimensions with holonomy
$\Sp(1)\ltimes \RR^4$ has a parallel null vector, and hence it is a
gravitational $pp$-wave of the type studied by Brinkmann in the 1920s
\cite{Brinkmann2}, but one which in addition admits a parallel spinor.
In adapted coordinates $x^+,x^-,x^i$ with $i,j=1,\dots,4$ such that
the parallel vector is $\d_-$, the metric has the following form
\cite{JMWaves}:
\begin{equation*}
  ds^2 = 2 dx^+ dx^- + a (dx^+)^2 + b_i dx^i dx^+ + g_{ij} dx^i dx^j~,
\end{equation*}
where $a$, $b_i$ and $g_{ij}$ are independent of $x^-$ and
$g_{ij}(x^+,x^i)$ is a $x^+$-dependent family of four-dimensional
hyperkähler metrics.  It is possible to change coordinates so that the
mixed term $b_i dx^i dx^+$ is absent, and we will do this.  Moreover we
will assume for simplicity of exposition that the $x^+$-dependence of
$g_{ij}$ is via a conformal factor.  The metric we will consider will
therefore be given by
\begin{equation*}
  ds^2 = 2 dx^+ dx^- + a(x^+,x^i) (dx^+)^2 + e^{2\sigma(x^+)}
  g_{ij}(x^i) dx^i dx^j~,
\end{equation*}
with $g_{ij}$ now a fixed hyperkähler metric.

The only nonzero component of the Ricci tensor is given by
\begin{equation*}
  R_{++} = -\half \bigtriangleup a - 4 \left(\sigma'' +
    (\sigma')^2\right)~,
\end{equation*}
where the ${}'$ denotes the derivative with respect to $x^+$ and
$\bigtriangleup$ is the laplacian relative to the metric $g_{ij}$.
If we let $a(x^+,x^i) = -8 \left(\sigma'' + (\sigma')^2\right) f(x^i)$,
then the Ricci curvature vanishes provided $f$ obeys $\bigtriangleup f =
1$.  Such an $f$ can always be found, at least locally.

As an example, we can consider $g_{ij}$ to be flat, and $f = \frac18
x^2$.  This gives a new family of $\nu=\frac14$ solutions, labelled by
the function $\sigma$.

\section{More Ricci-flat domain walls}

In this section we discuss generalised domain walls, focusing on the
$\D8$-brane in the massive IIA supergravity theory of Romans
\cite{Romans-Massive} for concreteness; although it should be clear
from this discussion how to generalise this to other domain walls, as
discussed in \cite{BP}.

The background of the $\D8$-brane \cite{PolchinskiWitten,BdRGPT} is
given by:
\begin{equation*}
  ds^2 = H^{\frac18} ds^2(\EE^{8,1}) + H^{\frac98} dy^2
  \qquad\text{and}\qquad e^\phi = H^{-\frac54}
\end{equation*}
with all other fields set to zero, and where $H(y)$ is given by
\begin{equation*}
  H(y) = 1 + m |y-y_0|~,
\end{equation*}
with $m$ the mass parameter in the supergravity theory.  This metric
has a delta-function singularity at $y=y_0$, which can be cured by
adding a source for the brane.  This solution preserves a fraction
$\nu=\half$ of the supersymmetry.

Following \cite{BP}, we define a new variable $z$ by
\begin{equation*}
  dz^2 = H^{\frac98} dy^2~,
\end{equation*}
so that the metric becomes
\begin{equation*}
  ds^2 = \left( 1 + \tfrac{25}{16} m |z-z_0| \right)^{\frac2{25}}
  ds^2(\EE^{8,1}) + dz^2,
\end{equation*}
and
\begin{equation*}
  e^\phi = \left( 1 + \tfrac{25}{16} m |z-z_0| \right)^{-\frac45}~.
\end{equation*}
This expression suggests the following generalised $\D8$-brane Ansatz
\cite{BP}:
\begin{equation*}
  ds^2 = F^2(z) ds^2(X) + dz^2~,
\end{equation*}
where $X$ is a nine-dimensional lorentzian manifold and $F$ and $\phi$ 
are functions of the real variable $z$.  As shown in \cite{BP}, the
choice
\begin{equation*}
  F(z) = \left( 1 + \tfrac{25}{16} m |z-z_0| \right)^{\frac1{25}}
  \qquad\text{and}\qquad e^\phi = F^{-20}~,
\end{equation*}
gives rise to a solution of the supergravity theory provided that $X$
is Ricci-flat.  In addition the solution preserves some supersymmetry
provided that $X$ admits \emph{parallel} spinors.

The case of $ds^2(X) = -dt^2 + ds^2(Y)$, with $Y$ an eight-dimensional
riemannian manifold of holonomy contained in $\Spin(7)$ was discussed
in detail in \cite{BP}.  However, according to
Table~\ref{tab:spinisotropy}, we see that there are also solutions
corresponding to indecomposable manifolds $Y$ with holonomy contained
in $G_2\ltimes \RR^7$, provided that they are Ricci-flat.

A possible construction of such metrics in the following.  Choose
coordinates $(x^+,x^-,x^i)$, with $i=1,\dots,7$, and consider the
following metric
\begin{equation*}
  ds^2(Y) = 2 dx^+ dx^- + a(x^+,x^i) (dx^+)^2 + e^{2\sigma(x^+)}
  g_{ij} dx^i dx^j~,
\end{equation*}
where $g$ is a metric of holonomy contained in $G_2$.  The nonzero
component of the Ricci tensor is given by
\begin{equation*}
  R_{++} = -\half \bigtriangleup a - 7 \left(\sigma'' +
    (\sigma')^2\right)~,
\end{equation*}
whence if we let $a(x^+,x^i) = - 14 \left(\sigma'' +
  (\sigma')^2\right) f(x^i)$, then the metric is Ricci-flat provided
that $\bigtriangleup f = 1$, which always has a local solution.

For example, if we take $g$ to be flat, then we can choose $f=\half
x^2$, and obtain a family of Ricci-flat metrics parametrised by the
function $\sigma$.  This family preserves a fraction $\nu=\frac14$ of
the supersymmetry.  Other fractions are possible by allowing $g$ to
have larger holonomy; for example: $\nu=\frac18$ if $g$ has holonomy
$\Sp(1)$, $\nu=\frac1{16}$ if $g$ has holonomy $\SU(3)$, and
$\nu=\frac1{32}$ if $g$ has holonomy $G_2$.

\section{Ricci-flat $\D3$-branes}

In the previous two sections we have seen how to generalise the examples
of Ricci-flat branes obtained in \cite{BP} to nonstatic brane
worldvolumes.  In this section we show how to curve the worldvolume of a
$\D3$-brane.  This case was not covered in \cite{BP} because it is not
possible to curve Ricci-flat static worldvolume metrics.  Indeed, if a
four-dimensional Ricci-flat lorentzian metric is decomposable, then it
is automatically flat.

The metric for the $\D3$-brane of type IIB is given by
\cite{HorowitzStrominger},
\begin{equation*}
g = H^{-\frac12}\, ds^2(\EE^{3,1}) + H^{\frac12}\, ds^2(\EE^6)~,
\end{equation*}
where now $\EE^{3,1}$ is the four-dimensional minkowskian worldvolume of
the threebrane and $\EE^6$ is the euclidean transverse space.  The
self-dual $5$-form $F$ has (quantised) flux on the unit transverse
five-sphere $S^5\subset \EE^6$ and the dilaton is constant.  $H$ is
again harmonic in $\EE^6$ and the unique spherically symmetric solution
with $\lim_{r\to\infty} H(r) =1$ is
\begin{equation*}
H(r) = 1 + \frac{a^4}{r^4}~; \qquad a^4 \equiv 4\pi g N \ell_s^4
\end{equation*}
where $g$ is the string coupling constant, given by the exponential of
the constant dilaton, and $\ell_s = \sqrt{\alpha'}$ is the string
length.  The solution corresponds to $N$ parallel $\D3$-branes at $r=0$.
The ten-dimensional Planck length is $\ell_p \equiv g^{\tfrac14}
\ell_s$, so that $a$ can be rewritten as
\begin{equation*}
a^4 =  4\pi N \ell_p^4~.
\end{equation*}

Just like was done in the previous two sections, we can try to curve the 
worldvolume metric while preserving supersymmetry and satisfying the
supergravity equations of motion.  As was explained above, the only
possibility is to consider an indecomposable four-dimensional metric.
According to Table~\ref{tab:spinisotropy}, for the metric to admit a
parallel spinor, it must have holonomy $\RR^2 \subset \Spin(3,1)$.  The
most general local metric with this holonomy was constructed, for
example, in Section~3.2.3 in \cite{JMWaves}, and it can be written as
follows:
\begin{equation*}
  ds^2 = 2 dx^+ dx^- + a (dx^+)^2 + b \epsilon_{ij} x^j dx^i dx^+ +
  dx^i dx^i~,
\end{equation*}
where $b=b(x^+)$ and $a=a(x^+,x^i)$ are arbitrary functions, and where
now $i,j=1,2$.  The only nonzero component of the Ricci tensor is
again
\begin{equation*}
  R_{++} = -\half \bigtriangleup a + 2 b^2~,
\end{equation*}
where $\bigtriangleup = \d_i\d_i$ is the laplacian on functions of the
transverse coordinates $x^i$.  In order for this to cancel, we need that 
$\bigtriangleup a$ should be a non-negative function of only $x^+$, so
that we can choose $b = \half \left(\bigtriangleup a\right)^{\half}$.
The most general local solution is given by
\begin{equation*}
  a = f(x^+) + g_i(x^+) x^i + e^{2h(x^+)} x^2 \qquad\text{and}\qquad
  b= e^{h(x^+)}~,
\end{equation*}
where the four functions $f$, $g_i$, and $h$ are arbitrary.
Such a $\D3$-brane solution preserves a fraction $\nu=\frac14$ of the
supersymmetry.

Interestingly enough, we can also deform the transverse space to the
threebrane.  One way to do this is to notice that the euclidean metric
in $\EE^6$ is that of the metric cone over the unit five-sphere:
\begin{equation*}
  ds^2(\EE^6) = dr^2 + r^2 ds^2(S^5)~.
\end{equation*}
As has been explained, for example, in \cite{AFHS}, one can substitute
the five-sphere for any five-dimensional Sasaki--Einstein manifold,
and recover a solution with presumably one fourth of the supersymmetry
of the spherical solution.  In other words, we can construct a large
class of $\D3$-brane solutions which preserve a mere fraction
$\nu=\frac{1}{16}$ of the supersymmetry.\footnote{In fact, the
  fraction is at least $\frac1{16}$ but might actually be larger,
  since in this case not all the ``supergravity Killing spinors''
  (i.e., zero modes of the gravitino shift equation) need be given
  simply as tensor products of Killing spinors in the two factors of
  the near horizon geometry.}

It is possible to analyse the near-horizon geometry of these
threebrane solutions and to study the Lie superalgebra of symmetries
as in \cite{JMFKilling}.  Since the near-horizon geometry does not
contain an anti de~Sitter factor, the putative Maldacena dual field
theory is not conformal, and in fact is not even Lorentz invariant.
This is consistent with the heuristic picture of understanding the
Maldacena dual as living on the brane, which is now a curved manifold.

%

\begin{thebibliography}{10}

\bibitem{AFHS}
BS~Acharya, JM~Figueroa-O'Farrill, CM~Hull, and B~Spence, \emph{Branes at
  conical singularities and their dual field theories}, Adv. Theor. Math. Phys.
  \textbf{2} (1998), 1249--1286, \texttt{hep-th/9808014}.

\bibitem{BdRGPT}
E~Bergshoeff, M~de~Roo, MB~Green, G~Papadopoulos, and PK~Townsend,
  \emph{Duality of type {II} 7 branes and 8 branes}, Nucl. Phys. \textbf{B470}
  (1996), 113--135, \texttt{hep-th/9601150}.

\bibitem{BP}
DR~Brecher and MJ~Perry, \emph{Ricci-flat branes}, \texttt{hep-th/9908018}.

\bibitem{Brinkmann2}
HW~Brinkmann, \emph{Einstein spaces which are mapped conformally on each
  other}, Math. Ann. \textbf{94} (1925), 119--145.

\bibitem{JMWaves}
JM~Figueroa-O'Farrill, \emph{Breaking the {$\M$}-waves},
  \texttt{hep-th/9904124}.

\bibitem{JMFKilling}
\bysame, \emph{On the supersymmetries of {A}nti-de~{S}itter vacua}, Class.
  Quant. Grav. (1999), no.~16, 2043--2055, \texttt{hep-th/9902066}.

\bibitem{FSDNW}
JM~Figueroa-O'Farrill and S~Stanciu, \emph{More {$\D$}-branes in the
  {N}appi--{W}itten background}, \texttt{hep-th/9909164}.

\bibitem{Guven}
R~Güven, \emph{Black $p$-brane solutions of {$D{=}11$} supergravity theory},
  Phys. Lett. \textbf{276B} (1992), 49--55.

\bibitem{HorowitzStrominger}
GT~Horowitz and A~Strominger, \emph{Black strings and $p$-branes}, Nucl. Phys.
  \textbf{B360} (1991), 197--209.

\bibitem{Janssen}
B~Janssen, \emph{Curved branes and cosmological $(a,b)$-models},
  \texttt{hep-th/9910077}.

\bibitem{PolchinskiWitten}
J~Polchinski and E~Witten, \emph{Evidence for heterotic - type {I} string
  duality}, Nucl. Phys. \textbf{B460} (1996), 525--540,
  \texttt{hep-th/9510169}.

\bibitem{Romans-Massive}
LJ~Romans, \emph{Massive {N=2a} supergravity in ten-dimensions}, Phys. Lett.
  \textbf{169B} (1986), 374.

\bibitem{Wang}
MY~Wang, \emph{Parallel spinors and parallel forms}, Ann. Global Anal. Geom.
  \textbf{7} (1989), no.~1, 59--68.

\end{thebibliography}
%

\providecommand{\bysame}{\leavevmode\hbox to3em{\hrulefill}\thinspace}

\end{document}